\documentclass[aps,pra,twocolumn,showpacs,floatfix]{revtex4}
\usepackage{graphicx,amsfonts,amsmath,bm}

\newcommand{\z}{{\bf z}}
\newcommand{\x}{{\bf x}}
\newcommand{\p}{{\bf p}}
\newcommand{\q}{{\bf q}}
\newcommand{\vv}{{\bf v}}
\newcommand{\g}{\gamma}

\begin{document}

\title{Time propagation of constrained coupled Gaussian wave packets}
\author{Toma\v{z} Fab\v ci\v c}
\author{J\"org Main}
\author{G\"unter Wunner}
\affiliation{Institut f\"ur Theoretische Physik 1, Universit\"at Stuttgart,
 70550 Stuttgart, Germany}
\date{\today}

\begin{abstract}
The dynamics of quantum systems can be approximated by the time propagation
of Gaussian wave packets.
Applying a time dependent variational principle, 
the time evolution of the parameters of the coupled Gaussian wave packets can be
calculated from a set of ordinary differential equations.
Unfortunately, the set of equations is ill-behaved in most practical
applications, depending on the number of propagated Gaussian wave packets,
and methods for regularization are needed.
We present a general method for regularization based on applying adequate nonholonomic
inequality 
constraints to the evolution of the parameters, keeping the equations of
motion well-behaved.
The power of the method is demonstrated for a non-integrable system with
two degrees of freedom.

\end{abstract}

\pacs{03.65.-w, 04.30.Nk}


\maketitle

\section{Introduction}
The method of Gaussian wave packet propagation is a popular tool for
quantum dynamics computations. Within this approximation it is assumed that
an initially Gaussian wave packet (GWP) stays Gaussian for all times.
The time evolution of the wave packet is given by the
time evolution of its parameters like width, phase, center, and momentum \cite{Hel75}. 
For a single GWP, this rather crude approximation is in general only valid 
for short time propagation. The approximation can be significantly improved, if a
 superposition of GWP is used and these GWP are propagated in
concert, since the number of adjustable parameters is increased and the
overall wave function is no longer restricted to a Gaussian shape \cite{Hel76,Saw85,Hea86,Han89}.
The equations of motion for the Gaussian parameters are obtained from a
time dependent variational principle (TDVP). It is well known that these  
coupled equations of motion for the time dependent parameters become
ill-conditioned 
from time to time during the
integration depending on how many GWP are used. The reasons for the ill-conditioned behavior of the
differential equations are near singularities of a matrix
 that has to be inverted after each time step of integration
\cite{Sko84,Saw85,Hea86,Han89,Kay89}.
Using step size control the time steps of the integration algorithm can become 
extremely small making
the method impracticably slow.  
In the worst case even a failure of the numerical matrix inversion or the 
further integration may occur. 

Different solutions to this
numerical problem were proposed, e.g.\ a regularization based on a 
singular value decomposition \cite{Kay89}.
The singular value decomposition is capable of 
regularizing the equations of motion in the sense that the method does not 
break down, however it does not solve the problem with the tiny 
step sizes \cite{Fab07}.
Another proposal is to adjust the number of GWP during evolution 
by increasing or reducing their number depending on whether the wave
function spreads or shrinks 
to avoid redundancy \cite{Saw85,Hea86,Hor04}. 

 It
has also been discussed to simplify the equations of motion by
keeping the widths of the propagated GWP fixed, called frozen Gaussian
approximation \cite{Hel81,Saw85,Han89,Hea86}, 
or much cruder, to neglect the
coupling between the GWP \cite{Saw85,Han89}. 
Another proposal is to reduce the variational freedom by forcing the GWP
to run on their classical trajectories \cite{Hel76,Sko84,Zop05}.
 But of course these grave restrictions
 severely reduce the accuracy of the GWP method.

Here we present a novel method to overcome the numerical problems or more
precisely a method that avoids numerical problems in the first place.
The idea is to impose adequate nonholonomic inequality constraints to 
the motion of each GWP, keeping the matrix regular. These constraints only
become active when it is numerically necessary and otherwise leave the full
variational freedom of the trial function.   
The method presented here is general and allows for the application of arbitrary 
(inequality) constraints not only on GWP trial functions.
There is numerical evidence, that near matrix
singularities usually result from widely varying amplitudes of largely
overlapping GWP. 
In our calculations it was sufficient to account for one ingredient of
the matrix singularity only, i.e.\ to
constrain the amplitudes of the individual GWP
to a reasonable domain. 
We account for the constraints in the time dependent variational
principle and obtain different equations of motion as compared to
the unconstrained variation. However, the equations of motion still have the
form of a matrix equation as in the unconstrained case.  
Properly chosen constraints only slightly decrease the accuracy of 
the variational approximation.
The additional error introduced by the constraints decreases with
a growing number of GWP.
The method is able to avoid numerical problems rendering the 
integration by orders of magnitude faster. 

The article is organized as follows. In section \ref{TDVP} we recapitulate the
time dependent variational principle. The equations of motion for the
Gaussian parameters obtained
from the TDVP applied to GWP are given for completeness. In section \ref{TDVPwithCon}
 we account for the inequality constraints in the TDVP and derive the regularized
 equations of motion. In section \ref{numRes} we compare numerical results obtained from
 the GWP method with and without constraints
 in a two-dimensional non-integrable model potential, namely the
 2D diamagnetic hydrogen atom. The accuracy of the
 constrained method is demonstrated by comparison with other propagation techniques.
 A summary is given in section \ref{summary}.

\section{Time dependent variational principle} \label{TDVP}
 The evolution of a quantum mechanical wave function is determined by
the Schr\"odinger equation
\begin{equation*}
i\dot{\psi}(t)=H \psi(t) 
\end{equation*}
where the wave function $\psi(t)$ is an element of the Hilbert space. 
An approximate solution $\chi(t)$ on a given manifold in Hilbert space 
can be obtained by a TDVP \cite{Dir30,Fre34,McL64,Kra83}. Here we choose
the formulation of McLachlan \cite{McL64}, or
equivalently the minimum error method \cite{Saw85}, where the norm of the deviation between
the right and
the left hand side of the Schr\"odinger equation with respect to the trial function
 is to be minimized. The quantity 
\begin{equation*}
I = ||i \phi(t) -H \chi(t)||^2 \overset{!}{=}{\rm min}
\end{equation*} 
is to be varied with respect to $\phi$ only, and then $\dot
\chi\equiv \phi$ is chosen.
We assume the approximation manifold to be parametrized by a set
of time dependent parameters $\z(t)=(z_1(t),\dots,z_{n_p}(t))$, i.e.\ $\chi(t) =\chi(\z(t))$. 
In terms of these parameters the quantity $I$ reads
\begin{eqnarray}
I & = &\left\langle \frac{\partial \chi}{\partial \z}\cdot \dot \z\Big|
\frac{\partial \chi}{\partial \z}\cdot \dot \z \right\rangle
-i\left\langle H\chi\Big|\frac{\partial \chi}{\partial \z}\cdot \dot \z \right\rangle \nonumber \\
& & +i\left\langle \frac{\partial \chi}{\partial \z}\cdot \dot \z \Big|H\chi \right\rangle
+\left\langle H\chi\Big|H\chi\right\rangle \label{errfuncpar}
\end{eqnarray}
which is a quadratic function of $\dot
\z$ for fixed values of $\z$. 
The variation $\delta \phi$ carries over to variations $\delta \dot \z$
leading to the condition 
\begin{equation}
\frac{ \partial I }{ \partial
 \dot z_{j}}=0, \quad j=1,\dots,n_{p}.
\end{equation}
 For complex parameters $z_j = z_{jr}+iz_{ji}$ one has the freedom to
take either $\partial I / \partial \dot z_{jr}  =0$ and $\partial I / \partial \dot z_{ji} =0$ or
to treat $\dot z_{j}^*$ and $\dot z_{j}$ formally as independent parameters
and to take either $ \partial I / \partial
\dot z_{j} = 0 $ or  $ \partial I / \partial
\dot z_{j}^* = 0 $. The resulting equations of motion are equivalent and read 
 \begin{equation}
 K \dot{\z}= -i {\bf h} \label{gg}
\end{equation}
in case of complex parameters $\z$, where
\begin{equation} 
K=\left\langle \frac{\partial \chi}{ \partial{\z}}
\Big|\frac{\partial \chi}{\partial{\z}} \right\rangle, \quad {\bf
  h}=\left\langle \frac{\partial \chi}
{ \partial{\z}}
\Big|H \Big|\chi \right\rangle.
\end{equation}
The Hermitian matrix $K$ is positive semi-definite
since 
\begin{equation}
{\bf c}^\dagger \left\langle \frac{\partial \chi}{\partial \z} \Big|
\frac{\partial \chi}{\partial \z}\right\rangle {\bf c} =\left\langle
\frac{\partial \chi}{\partial \z}\cdot {\bf c} \Big|
\frac{\partial \chi}{\partial \z}\cdot {\bf c}
\right\rangle=\Big|\Big|\frac{\partial \chi}{\partial \z}\cdot {\bf c} \Big|\Big|^2\geq 0, 
\end{equation}
$\forall\; {\bf c} \in {\mathbb C}^{n_p}$, ensuring that
the extremum of the quadratic quantity $I$ is a minimum.

The Schr\"odinger equation is replaced by a system of ordinary first order
differential equations of motion for the parameters $\z(t)$ where after every time
step of integration the set
of simultaneous linear equations (\ref{gg}) must be solved for the time
derivatives $\dot \z$ if a numerical algorithm for ordinary differential equations,
e.g.\ Runge-Kutta or Adams, is used.

\subsection{Application of the TDVP to GWP}
In this article a superposition of GWP as trial function is discussed.
Each GWP ($\x \in \mathbb{ R}^D$) is of the form 
\begin{equation}
g({\bf y}^k,\x)=e^{i((\x-\q^k)A^k(\x-\q^k)+\p^k\cdot(\x-\q^k)+\gamma^k)}, 
\end{equation}
where $A^k$ is a complex symmetric $D\times D$ matrix, the momenta $\p^k$
and centers $\q^k$ are real, $D$-dimensional vectors, and the phase and normalization are given by the
complex scalars $\g^k$. The Gaussian parameters of the $k$-th GWP are denoted
by ${\bf  y}^k=(A^k,\p^k,\q^k,\g^k)$. Their time argument is omitted for
brevity. The trial function
is a superposition of $N$ such GWP 
\begin{equation}
\chi(\z,\x)=\sum_{k=1}^N g({\bf y}^k,\x), \quad \z=({\bf y}^1,\dots ,{\bf
  y}^N) \label{chi} .
\end{equation}
Using a splitting of the Hamiltonian
$H=T+V$ we obtain
\begin{eqnarray}
i \dot \chi-T \chi & = & 
\sum_{k=1}^N g({\bf y}^k,\x)([i{\rm tr}\,A^k-\dot \g^k+\p^k\cdot(\dot
\q^k-\frac{1}{2}\p^k)]\nonumber \\ 
& & \quad +[-\dot\p^k+2 A^k(\dot  \q^k-\p^k)]\cdot(\x-\q^k)\nonumber \\ 
& & \quad +(\x-\q^k)[-\dot A^k-2(A^k)^2](\x-\q^k))\nonumber \\ 
& \equiv & \sum_{k=1}^N(v_0^k+\vv_1^k\cdot\x+\frac{1}{2}\x V_2^k \x)\,g({\bf y}^k,\x),
\label{v_koef_def}
\end{eqnarray}
which defines, after sorting by powers of $\x$, the complex scalars
$v_0^k$, the complex vectors $\vv^k_1 \in {\mathbb C}^D$ and the
complex symmetric $D \times D$ matrices $V_2^k$ as the coefficients of a
second order polynomial.
According to the TDVP these coefficients $(v^k_0,\vv^k_1,V^k_2),\; k=1,\dots,N$ are
calculated from a set of linear equations
\begin{gather}
\sum_{k=1}^N  v_0^k \langle g^l|x^m_ix^n_j|g^k\rangle + \sum_{k=1}^N \langle g^l|x^m_ix^n_j
\x \cdot {\bf v}_1^k  | g^k\rangle  \nonumber  \\ 
+\frac{1}{2} \sum_{k=1}^N \langle g^l|x^m_ix^n_j
\x V_2^k \x  | g^k\rangle    
= \sum_{k=1}^N \langle g^l|x^m_ix^n_j V(\x)|g^k
\rangle; \label{linGl}
\end{gather}
\begin{equation*}
l=1,\dots ,N; \quad    m+n =0,1,2;  \quad i,j=1,\dots,D.  
\end{equation*}
On the right hand side the potential $V(\x)$ of the Hamiltonian is
inserted.
It is straightforward to calculate the time
derivatives of the Gaussian parameters once the linear equations (\ref{linGl})
are solved, since the differential equations for
the Gaussian parameters can be expressed by $(v^k_0,\vv^k_1,V^k_2),\;
k=1,\dots,N$ according to their definition in equation (\ref{v_koef_def}):
\begin{equation}
\begin{array}{ccl}
\dot{A}^k & = & -2 (A^k)^2-\frac{1}{2} V_2^k, \\[5pt]
\dot{\q}^k & = & \p^k+{\bf s}^k,  \\              [5pt]    
\dot \p^k  & = & 2{\rm Re}\,A^k {\bf s}^k-{\rm Re}\,{\bf v}_1^k-{\rm Re}\,V_2^k \q^k, \\[5pt]
\dot{\g}^k & = & -v_0^k+i{\rm tr}\, A^k+ \frac{1}{2}(\p^k)^2-{\bf  v}_1^k\cdot \q^k \\[5pt]
           &   & -\frac{1}{2} \q^k V_2^k \q^k+\p^k\cdot{\bf s}^k,
\end{array}\label{ggg} 
\end{equation}
where ${\bf s}^k = 
\frac{1}{2}({\rm Im}\,A^k)^{-1}({\rm Im}\,{\bf v}_1^k + {\rm Im}\, V_2^k
\q^k)$.
Numerically it is more appropriate to introduce two
additional $D\times D$ complex matrices $B^k,C^k$ according to
$A^k=\frac{1}{2}B^k(C^k)^{-1}$, and to integrate the
equations of motion 
\begin{equation}
\begin{array}{ccl}
\dot C^k  & = & B^k, \\
\dot B^k & = & -V_2^k C^k
\end{array}
\end{equation} 
instead of integrating $A^k(t)$ directly, because the
oscillating $(A^k(t))^2$ term causes numerical difficulties \cite{Hel76a}. 
For numerical accuracy, it is appropriate to symmetrize the matrix $A^k(t)$ 
after each time step.

Equation (\ref{linGl}) can be abbreviated by $K \vv ={\bf r}$
  when all coefficients $(v^k_0,\vv^k_1,V_2^k),\; k=1,\dots,N$ are put
  together into the complex vector $\vv$. 
All inner products in Hilbert space denoted by $\langle .|.\rangle$ are
calculated in position space representation. The integrals that build up the
components of the matrix $K$ on the left hand side of equation
(\ref{linGl}) as well as the integrals 
on the right hand side can be solved analytically, provided the potential 
is of special form, e.g.\ polynomial, Gaussian or exponential.

Given some initial wave function, i.e.\ the initial parameters $\z(t=0)$,
 the wave function is propagated by integrating the
trajectories of the parameters. At every time step equation (\ref{linGl}) must be
solved for the coefficients $\vv$ which are inserted in 
(\ref{ggg}) to obtain $\dot \z$.
In the course of integration, depending on how many GWP are propagated
in common, it will sooner or later happen that the matrix $K$ associated
with the set of linear equations (\ref{linGl}) becomes ill-conditioned, or even
numerically singular.
As a result the time step of the integration routine becomes extremely
small, rendering the method of GWP propagation impracticably slow. In the
worst case, further integration or matrix inversion respectively, can even fail. 

\section{Inequality constrained TDVP} \label{TDVPwithCon}  
Matrix singularity problems arise from overcrowding the basis set,
i.e.\ from situations where fewer GWP would be sufficient to represent the wave
function. On the other hand 
for an accurate approximation of the wave function
it is desirable to have a large number of
adjustable parameters.
However, there is a discrepancy between the number of
GWP necessary to give accurate results and the maximum number
of GWP that can be propagated using the TDVP without numerical difficulties
\cite{Han89}.
 As mentioned above there exist different proposals to overcome this
numerical problem, such as a singular value decomposition of the matrix
$K$ \cite{Kay89} or reducing the number of GWP when overcrowding takes
place \cite{Saw85,Hea86,Hor04}. Also reducing the variational freedom
by freezing the widths \cite{Hel81,Saw85,Han89,Hea86} and choosing
classical trajectories for the centers of the GWP \cite{Hel76,Sko84,Zop05} has been discussed.

Our approach of
regularizing the equations of motion for the parameters is
based on minimizing the quantity $I$ in (\ref{errfuncpar}) while certain
inequality constraints
are applied. The constraints must be chosen in such a way that they
prevent the matrix $K$ in (\ref{linGl}) to become  ill-conditioned.
This means all Gaussian parameters evolve freely according to the TDVP, and
 the constraints only
become active from time to time whenever the unconstrained evolution would drive the
parameters in domains where the matrix would be too singular, and
are switched off as soon as these 'forbidden' domains are left again.  
Formally spoken we reduce the space of admissible configurations to regions
where the associated matrix $K$ is regular. 

To demonstrate the generality of our method we first apply constraints to the general
case of an arbitrary trial function $\chi(\z(t))$ whose parameters
$\z(t)$ evolve according to equation (\ref{gg}). 
We derive their modified equations of motion which are obtained if    
the parameters $\z(t)$ are subject to some arbitrary inequality
constraints.
Then we return to GWP trial functions (\ref{chi})  and derive the modification of
equation (\ref{linGl}) obtained when the GWP are subject to inequality
constraints. Adequate constraints which prevent the matrix from
singularity are presented and applied.  

Due to real inequality constraints it is convenient to use a real formulation of the equations.
Complex quantities are split into their real and imaginary parts,
which are denoted by the subscripts $r$ and $i$, respectively.

\subsection{Inequality constrained TDVP on arbitrary trial functions}
Consider an arbitrary trial function $\chi(\z(t))$ and
assume a real inequality constraint on the parameters $\z(t) \in {\mathbb C}^{n_p}$ which can be
written in the form
\begin{equation}
f(\z,\z^*) \equiv f(\z_r,\z_i) \equiv f( \bar \z)\geq f_{\rm min} \label{const}
\end{equation}
where the function $f$ is explicitly known.
For brevity, the notation $ \bar \z \equiv ( \z_r, \z_i) \in {\mathbb R}^{2n_p}$
will be used. 

\begin{figure}
\includegraphics[width =.9\columnwidth]{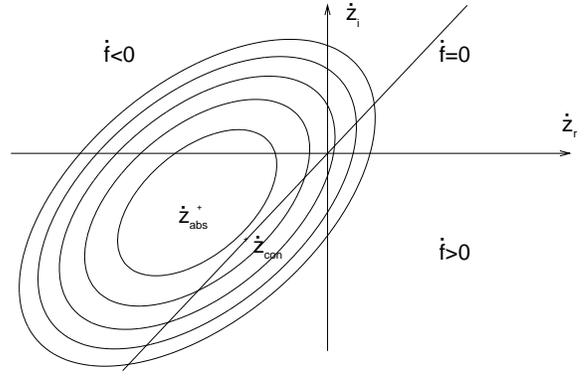}
\caption{The ellipses schematically represent isolines of 
$I$ in equation (\ref{errfuncpar}) for fixed parameters $\z$. The domain of allowed 
$\dot \z$ for the minimum of $I$ is the full space, when $f>f_{\rm min}$ and is reduced to the
half space $\dot f \ge 0$, when $f=f_{\rm min}$ is reached.}\label{iso}
\end{figure}

As long as $f(\z_r,\z_i)>f_{\rm min} $, all parameters evolve according to equation
(\ref{gg}) without being affected by the restriction.
When
$f(\z_r,\z_i) = f_{\rm min}$ is reached at some point in time $t$,
  the constraint becomes active, and we have to
demand $\dot f(t)\geq 0$, otherwise $f(t+\Delta t)$ with some small
positive $\Delta t$ would
violate the constraint (\ref{const}). Therefore the quantity $I$ of equation
(\ref{errfuncpar}) at fixed $\z$ must
be minimized with respect to $\dot \z$, where $(\dot \z_r, \dot \z_i)$ are now
subject to 
the constraint 
\begin{equation}
\dot f= \frac{\partial f}{\partial \z_r}\cdot \dot
\z_r+\frac{\partial f}{\partial \z_i}\cdot \dot \z_i 
\equiv \frac{\partial f}{\partial \bar \z}\cdot \dot{\bar \z} \geq 0. \label{linconst}
\end{equation}
In other words the possibly nonlinear constraint (\ref{const}) on $\z$ has
been reduced
 to the linear constraint (\ref{linconst}) on $\dot \z$ when
$f=f_{\rm min}$. Then 
the allowed domain of $(\dot \z_r,\dot \z_i)$ for searching the minimum of $I$ is no
more the whole space ${\mathbb R}^{2n_p}$, but
the half-space $\dot f \ge 0$ linearly restricted by equation (\ref{linconst}).
In general, minimization of a function on a given domain requires two steps,
firstly to find the local internal minima and secondly, to find the local minima on the boundaries. 
The global minimum in
the given domain is obtained by comparison. Here it is sufficient to
search for the minimum of $I$ solely on the boundary of the domain defined
by equation
(\ref{linconst}) where the equality sign is fulfilled. That means the inequality
(\ref{linconst}) may be replaced by the computationally much more feasible constraint 
 \begin{equation}
\frac{\partial f}{\partial \z_r}\cdot \dot
\z_r+\frac{\partial f}{\partial \z_i}\cdot \dot \z_i 
\equiv \frac{\partial f}{\partial \bar \z}\cdot \dot{\bar \z}= 0. \label{equalcon}
\end{equation}
 The reason is that 
 $I$ is a positive definite parabolic function of $\dot \z$ whose absolute
 minimum lies outside the allowed domain by assumption. Since there are no internal minima
  $I$ obviously takes its allowed minimum on the boundary 
of the allowed domain. 
The constraint is
 switched off again as soon as the trajectory $\dot \z(t)$ of
  the absolute minimum of $I$ crosses the plane given by equation
  (\ref{equalcon}) in the
$(\dot\z_r,\dot\z_i)$-space at fixed values of $(\z_r,\z_i)$. 
Note that arbitrary nonlinear constraints
 (\ref{const}) on $\z$ always lead to linear constraints (\ref{linconst}) on $\dot\z$
  leading to a linearly equality constrained quadratic minimization problem,
which can directly be solved by a matrix equation as in the
unconstrained case (\ref{gg}).
The strategy is
  illustrated in figure \ref{iso}, which shows schematically the elliptical
  isolines of $I$ for fixed $\z$ as a
  function of $(\dot \z_r,\dot \z_i)$. The values of the parameters $\z$
 determine the shape and the position of the parabola as well as the slope
 of the plane $\dot f =0$. 
In figure \ref{iso}, $\dot \z_{\rm abs}$ denotes the absolute minimum of $I$,
obtained from equation (\ref{gg}). The plane $\dot  f = 0$ (equation
(\ref{equalcon})) divides the
$2n_p$-dimensional $(\dot\z_r,\dot\z_i)$-space into the two half-spaces
$\dot  f < 0$ and $\dot  f \geq 0$. The point $\dot \z_{\rm con}$ is the constrained minimum
of $I$ 
in the half-space $\dot f \geq 0$, which lies on its boundary, i.e.\
on the plane  $\dot  f = 0$ as explained above.

  As long as $f>f_{\rm min}$, $ \dot \z_{\rm abs}$ determines the evolution of
  the parameters. 
However when $f=f_{\rm min}$ is reached, then $\dot \z_{\rm
    con}$ is taken for the further integration of the trajectories $\z(t)$
  until $\dot \z_{\rm abs}$,
 driven by the constrained evolution of the parameters, eventually crosses
 the plane $\dot  f = 0$ from $\dot  f < 0$ to $\dot  f > 0$ . 
At this point, $ \dot \z_{\rm abs}$ and $\dot \z_{\rm con}$ coincide
 and $\dot \z_{\rm abs}$ is taken again
  for further integration, since $\dot f>0$ leads to an increase
  of $f(t)$ with time, according to the constraint.

For the extension to multiple, say $m$, active constraints the real scalar valued 
function $f(\z_r,\z_i)$ is simply replaced by the real vector valued
function  ${\bf f}(\z_r,\z_i)\equiv {\bf f}(\bar \z)= (f_1,\dots,f_m) \in {\mathbb R}^m$. 

Now that the nonholonomic nonlinear inequality constraints (\ref{const}) on $\z$ are reduced to the
holonomic linear equality constraints (\ref{equalcon}) on $\dot \z$ by the constrained TDVP, we can 
determine the constrained minimum $\dot \z_{\rm con}$ by a standard method like Lagrangian multipliers. 
Alternatively, the constrained minimum can also be obtained by elimination of 
the dependent variational parameters. We prefer the method of Lagrange multipliers due to its 
generality.
The method of Lagrange multipliers yields a compact form of the equations of motion for arbitrary constraints
and the conditions for switching off the constraints are obtained with only little
additional numerical effort as will be shown below. 
Both methods however, require a minimization problem with equality constraints. When inequality constraints are 
applied, the elimination of dependent variational parameters is not possible.

We construct the function
\begin{equation}
L=I+{\boldsymbol \lambda} \bar M \dot{\bar \z}
\end{equation}
with the Lagrangian multipliers ${\boldsymbol \lambda} \in {\mathbb R}^m$
and the real valued $m \times 2n_p$ matrix 
$\bar M = \frac{\partial {\bf f}}{\partial \bar \z}$.
The minimum of $I$ under the
constraint (\ref{linconst}) is found by $\partial L / \partial {\boldsymbol
  \omega} = 0$
where 
 \begin{equation*}
{\boldsymbol \omega}\equiv\left(
\begin{array}{c}
 \dot \z_r \\
 \dot \z_i \\
{\boldsymbol \lambda}
\end{array}
\right)
\equiv 
\left(
\begin{array}{c}
\dot{\bar \z}\\
{\boldsymbol \lambda}
\end{array}
\right)
  \in {\mathbb R}^{2n_p+m}.
\end{equation*}
We obtain a set of linear equations  
\begin{equation}
\left(
\begin{array}{c|c}
\bar K & \bar M^T \\
\hline
\bar M  & 0
\end{array}
\right)
\left(\begin{array}{c} \dot{\bar \z} \\ {\boldsymbol \lambda}
\end{array} \right) 
= 
\left( \begin{array}{c} \bar{\bf h} \\ 0
\end{array} \right),\label{ggcon}
\end{equation}
with
\begin{equation}
\bar K =
\left(
\begin{array}{cc}
K_r & -K_i \\
K_i &  K_r \\
\end{array}
\right),\;
\bar{\bf h} 
=
\left(
\begin{array}{c}
 {\bf h}_i \\ 
 -{\bf h}_r
\end{array}
\right)
, 
\end{equation}
where the matrix $K$ and the vector ${\bf h}$ are the complex quantities of
equation (\ref{gg}).
 If no constraint is active, i.e.\ $m=0$, then equation
(\ref{ggcon}) obviously reduces to the real formulation of equation
(\ref{gg}).
We use a real formulation, i.e.\ complex quantities are split into
their real and imaginary parts, 
because real constraints like $f>f_{\rm min}$ naturally lead to
real Lagrangian multipliers. 

 The constraint (\ref{equalcon}) is switched off
again when $\dot \z_{\rm abs}$ crosses the plane $\dot f = 0$ from $\dot
f<0$ to $\dot f>0$. Finding this event can be accomplished in two
ways. The trivial but
computationally expensive way is to calculate
not only $\dot \z_{\rm con}$ from (\ref{ggcon}), which is needed for
integration, but
 additionally $\dot \z_{\rm
  abs}$ (from equation (\ref{gg})) after every time step of integration and to check when $\dot
f|_{\dot \z_{\rm abs}}$
changes its sign. This inefficient procedure would require the solution of a complex $n_p \times n_p$ 
matrix equation  for $\dot \z_{\rm abs}$ and additionally the solution of the real $(2n_p+m)\times(2n_p+m)$ matrix 
equation for $\dot \z_{\rm con}$.
However it is much more efficient to check when $\lambda$
changes its sign for the special case $m=1$.  
If more than one constraint is active, $m>1$, it is recommended to solve the matrix equation 
(\ref{ggcon}) by decomposition into two blocks, as indicated by 
the horizontal line in equation (\ref{ggcon}), namely  into 
\begin{equation}
\bar K \dot{\bar \z}+\bar M^T {\boldsymbol \lambda} =\bar{\bf h}\label{ggcon1}
\end{equation}
obtained by the upper part of equation (\ref{ggcon}),
and the lower part
\begin{equation}
\bar{M}\dot{\bar \z}=0,
 \label{ggcon2}
\end{equation}
which represents the active constraints.
The solution for the unknowns $\dot{\bar \z},{\boldsymbol \lambda}$ is obtained by first
solving equation (\ref{ggcon1}) for $\dot{\bar \z}$ 
\begin{equation}
\dot{\bar\z} = \bar{K}^{-1}\bar{\bf h}-\bar{K}^{-1}\bar{M}^T{\boldsymbol \lambda}\label{ggcon1a}
\end{equation}
and inserting it in equation (\ref{ggcon2}) 
in order to eliminate $\dot{\bar \z}$. 
The result is a small $m \times m$ matrix equation for determining ${\boldsymbol \lambda}$
\begin{equation}
\underbrace{
\bar{M} \bar{K}^{-1} \bar{M}^T
}_{m \times m}
{\boldsymbol \lambda} 
= \bar{M} \bar{K}^{-1}\bar{\bf h}\in {\mathbb R}^m. \label{ggcon3}
\end{equation}
The conditions for switching off any of the active constraints are now contained in 
the right hand side of equation (\ref{ggcon3}), since 
\begin{equation}
\dot{\bf f}|_{\dot \z_{\rm abs}} 
\equiv \frac{\partial {\bf f}}{\partial \bar \z} \dot{\bar \z}_{\rm abs}
\equiv \bar M \dot{ \bar \z}_{\rm abs}
\equiv  \bar{M} \bar{K}^{-1} \bar{\bf h}
\end{equation}
due to the definitions. The $i$th active constraint ($1 \le i \le m$) is to be switched off when the 
$i$th component of $\dot{\bf f}|_{\dot \z_{\rm abs}}$ changes its sign from minus to plus. 

When we insert the Lagrange multipliers calculated from (\ref{ggcon3}) in (\ref{ggcon1a}) we obtain 
$\dot \z_{\rm con}$, needed for propagation.
Numerically, the calculation of $\bar K^{-1}\bar{\bf h}$ and $\bar K^{-1}\bar M^T$ in (\ref{ggcon3})
requires only one factorization of the large matrix $\bar K$. After multiplying with
$\bar M$ from the left the small set of linear equations (\ref{ggcon3}) for determining 
$\boldsymbol{ \lambda}$ is obtained.
Compared to the factorization of $\bar{K}$ the solution of the 
$m \times m$ matrix equation (\ref{ggcon3}) for the Lagrange multipliers is negligible, since the number
 of parameters $n$ will in general exceed the number of constraints $m$ by far, e.g. in our numerical
calculation 
there is $2n_p=240$ and the number $m$ of simultaneously active constraints is not larger than three.


\subsection{Inequality constrained TDVP applied to GWP}
When GWP are used as trial function, it is convenient
 to formulate a  set of linear equations for the coefficients
 $\vv=\vv_r+i\vv_i$ first and then to obtain 
$\dot \z $ from (\ref{ggg}) in a second step, just as was done in 
 section \ref{TDVP}. 
For these coefficients $\vv_r$ and $\vv_i$, summarized by the notation
$(\vv_r,\vv_i)=\bar\vv$,
 a similar set of
linear equations is obtained. Equations (\ref{ggg}) which describe the
connection between the time derivatives of the parameters and the
coefficients, are written in real formulation, where all complex quantities
are split into their real and imaginary parts. We obtain 
\begin{equation}
\begin{array}{ccl}
\dot A^k_r  & = &                                  -\frac{1}{2}V^k_{2r}                                           -2((A^k_r)^2-(A^k_i)^2),  \\  [4pt]
\dot A^k_i  & = &                                                                                                  -\frac{1}{2}V^k_{2i}        -2A^k_r A^k_i-2A^k_iA_r^k      ,          \\ [4pt] 
\dot \p^k   & = &            -\vv^k_{1r}           -V^k_{2r}\q^k                +2A^k_r \Lambda^k \vv^k_{1i}  +2 A^k_r \Lambda^k V^k_{2i} \q^k ,  \\   [4pt]                         
\dot \q^k   & = &                                                                                          \Lambda^k \vv^k_{1i}  +\Lambda^k V^k_{2i} \q^k   +\p^k ,  \\  [4pt]
\dot \g^k_r & = & -v^k_{0r}  -\vv^k_{1r} \cdot\q^k  -\frac{1}{2}\q^k V^k_{2r}\q^k \\[4pt]
            &   &   + \p^k \Lambda^k \vv^k_{1i}  +\p^k \Lambda^k V^k_{2i} \q^k  -{\rm  tr}\,A^k_i+ \frac{1}{2}(\p^k)^2, \\[4pt]
\dot \g^k_i & = &                                                       
-v^k_{0i}   -\q^k \cdot \vv^k_{1i}  -\frac{1}{2} \q^k V^k_{2i} \q^k 
+{\rm tr}\,A^k_r  ,
\end{array} \label{glreal}
\end{equation}
with $\Lambda^k=\frac{1}{2}(A^k_i)^{-1}$. 

Using the notation $\bar
\z=(A^1_r,A^1_i,\p^1,\q^1,\g^1_r,\g^1_i,\dots,$ $A^N_r,A^N_i,\p^N,\q^N,\g^N_r,\g^N_i)$
the complete set of equation (\ref{glreal}) for all $k=1,\dots,N$, which are
linear in $(v_0^k,\vv_1^k,V_2^k)$, may be written in short form
$\dot{\bar{\z}} = \tilde{U} \bar\vv + \tilde{\bf d}$.
The matrix $\tilde U$ is block-diagonal with $N$ blocks.
Each block consists of those coefficients in equation (\ref{glreal}) linear
in $( v_0^k,\vv_1^k,V_2^k)$. The constant terms are absorbed in the
vector $\tilde{\bf d}$.
The linear equality constraint
(\ref{equalcon}) for a GWP trial function reads
\begin{equation}
\begin{array}{ccc}
\dot f & = &\sum_{k=1}^N\left( \frac{\partial f}{\partial A^k_r}\dot A^k_r
+  \frac{\partial f}{\partial A^k_i}\dot A^k_i 
+  \frac{\partial f}{\partial \p^k}\cdot \dot \p^k \right. \\
& &
\left. +  \frac{\partial f}{\partial \q^k}\cdot \dot \q^k
+  \frac{\partial f}{\partial \g^k_r} \dot \g^k_r
+  \frac{\partial f}{\partial \g^k_i} \dot \g^k_i
\right)= 0
\end{array}
 \label{GWPcon}
\end{equation}
where the notation 
\begin{equation}
\frac{\partial f}{\partial A^k_r}\dot A^k_r =
\sum_{l,j=1}^D \frac{\partial f}{\partial (A^k_r)_{lj}} (\dot
A^k_r)_{lj}
\end{equation}
is used. Expressing the time derivatives in equation (\ref{GWPcon})
 by the coefficients $\vv_r$ and $\vv_i$ using (\ref{glreal}),
 $m$ arbitrary constraints (${\bf f}=(f_1,...,f_m) \in {\mathbb R}^m$) imply
\begin{equation}
\dot {\bf f}= 
\frac{\partial {\bf f}}{\partial \bar\z}\tilde{U}\bar \vv 
+ \frac{\partial {\bf f}}{\partial \bar\z}\tilde{\bf d} \equiv
\bar U \bar \vv + \bar{\bf d} =0, \label{fdot1}
\end{equation}
and hence a set of linear equations for $(\vv_r,\vv_i)$ and the Lagrange
multipliers ${\boldsymbol \lambda}\in {\mathbb R}^m$ is obtained

 \begin{equation}
\left(
\begin{array}{c|c}
\bar K  & \bar U^T \\
\hline
\bar U    & 0
\end{array}
\right)
\left(\begin{array}{c}  \bar \vv \\{\boldsymbol \lambda}
\end{array} \right) 
= 
\left( \begin{array}{c} \bar{\bf r} \\ -\bar{\bf d}
\end{array} \right),
\end{equation}
with
\begin{equation}
\bar K = 
\left(
\begin{array}{cc}
K_r & -K_i \\
K_i & K_r
\end{array}
\right),\,
\bar{\bf r} =
\left(
\begin{array}{c}
{\bf r}_r \\
{\bf r}_i
\end{array}
\right)
. \label{gwp2}
\end{equation} 
Here, $K=K_r+iK_i$ and the vector ${\bf r}={\bf r}_r+i{\bf r}_i$ are the matrix
and the right hand side of equation (\ref{linGl}), respectively. 

We now have all equations needed for propagation of coupled GWP subject
to arbitrary 
constraints (\ref{const}). Instead of (\ref{linGl}) we solve (\ref{gwp2})
for $(\vv_r,\vv_i)$ (when no constraints are active both sets of equations
are equivalent) after each time step. These coefficients are inserted in
(\ref{ggg}) (or equivalently in (\ref{glreal})) to obtain the time
derivatives of the Gaussian parameters,
which are needed by the integration routine to integrate the next time step.  

In order to find convenient constraints it is necessary to investigate
the reasons for the numerical matrix singularity.
The generic reasons for an ill-conditioned matrix $K$
are twofold. One cause is a strong overlap of neighboring GWP, the
other cause is 
widely spread norms of the GWP.
A restriction on the norm of the GWP 
\begin{equation}
g_{\rm min}\leq ||g^k|| \leq g_{\rm max},\quad k=1,...,N \label{constr1}
\end{equation}
turns out to be sufficient to regularize the equations of motion.
 It is however more simple and numerically efficient
to impose the restrictions 
\begin{equation}
f_{\rm  min}\equiv \g_{\rm min}\leq f^k(\tilde \z) = {\rm Im}\g^k \leq
\g_{\rm max} \equiv f_{\rm  max}, \label{constr2}
\end{equation}
with $k=1,...,N$ on the amplitude of the GWP.
Both restrictions (\ref{constr1}) and (\ref{constr2}) are equivalent for
frozen GWP and they are similar even for thawed GWP (at least for
bounded systems where the width of the GWP is bounded by the potential).
For the  active constraints ($\g^k_i=\g_{\rm min}$ or $\g^k_i=\g_{\rm max}$) equation
({\ref{equalcon}) using (\ref{glreal}) translates into 
\begin{equation}
\dot \g^k_i = -v^k_{0i}-\q^k \cdot \vv^k_{1i} -\frac{1}{2} \q^k V^k_{2i}
\q^k +{\rm tr}\,A^k_r = 0 \label{imgpos}.
\end{equation}
Therefore, in the notation of equations (\ref{fdot1}) and (\ref{gwp2}) 
the entries of $\bar U$ are
mostly zero except for the terms of equation (\ref{imgpos}) and $\bar d={\rm tr}\, A^k_r$. 
This especially simple case of constraints, where Gaussian parameters are bounded directly, 
leads to simply temporary freezing these parameters $\g^k_i$
when $\g^k_i=\g_{\rm min}$ ($\g^k_i=\g_{\rm max}$) is reached.
As mentioned above, the equations of motion can instead of using Lagrange multipliers be 
alternatively obtained by elimination of the dependent parameters.
The frozen $\g^k_i$ must be simply ignored in the variation. 
However additional calculations are then necessary 
to find the criteria for switching off the constraints. 

Should in some cases the restriction on the amplitudes (\ref{constr2}) not
be adequate, an upper bound on the maximum of the allowed overlap of
neighboring GWP or a lower bound on the least eigenvalue of the matrix
may be applied.

\section{numerical results}\label{numRes}
Numerical tests using coupled GWP were often performed in one dimension,
e.g.\ on the Morse potential \cite{Saw85,Hea86,Han89}.  
Here we use a two dimensional non-integrable potential for testing our 
method. The Hamiltonian of the system represents the 
diamagnetic Kepler problem  in a 2D rotating $(x,z)$ frame (for review, see
e.g.\ \cite{Fri89,Has89}). The magnetic field axis is directed along the 
 $z$-axis.
The potential in regularized semiparabolic coordinates reads
\begin{equation}
V(\mu,\nu)=\alpha(\mu^2+\nu^2)+\frac{1}{8}\beta^2\mu^2\nu^2(\nu^2+\mu^2),
\end{equation}
with
\begin{equation}
r^2 = x^2+z^2, \quad
\mu^2 = r+z, \quad
\nu^2 = r-z.
\end{equation}
The parameters are set to $\alpha=1/2$ and $\beta=1/5$
in our calculations.

The method of free GWP propagation is compared to the method of
constrained GWP propagation.
The value of the lower bound in equation (\ref{constr2}) is $\g_{\rm min}= -6.5$,
an upper bound was not needed.
The comparison is presented in figure \ref{fig_div}. 
The trial wave function consists of eight GWP with the same initial values
for both calculations.
Solid lines represent results of the
free
propagation, dashed lines represent the results of constrained propagation.
In figure \ref{fig_div}(a) that normalization parameter $\g_i^{k}(t)$ is
selected and drawn that first 
reaches $\g_{\rm min}=-6.5$ at $t \approx 6.9\, t_{cl}$ where $t_{cl}$ is 
the classical period of small harmonic oscillations around the minimum of
the potential. This choice allows for a direct comparison, because the
trajectories of both calculations are equal before $\g_{\rm min}$ is reached for the first time
by any of the $\g^k$, and they differ afterwards. 
Normalization parameters of the other seven GWP are not
plotted but show similar qualitative behavior. In terms of
figure \ref{iso}, the trajectories $\g_i^k(t)$ in the range $6.9\, t_{cl} \lesssim t
\lesssim 7.15\, t_{cl}$ are obtained using $\dot \z_{\rm abs}$ for the
integration of the solid line 
 and using $\dot \z_{\rm con}$ for the integration of the dashed
 line. Obviously the trajectory represented by the dashed line
sticks to the value $\g_{\rm min}=-6.5$ till $t\approx 7.15 \,t_{cl}$ where $\dot \z_{\rm abs}$
crosses the plane $\dot \g_i^k=0$. This scenario repeats several times as can be seen
in the figure.  
Figure \ref{fig_div}(b)
compares the step sizes used by the variable step Adams routine to
integrate the trajectories. The integration of the unconstrained equations
of motion becomes extremely slow around
$t\approx 7.1 \, t_{cl}$, and later on again for several times where the
step sizes become tiny. Obviously there is a strong correlation between
very low values of $\g_i^k$ in panel (a) and extremely small step sizes in panel (b) 
for unconstrained propagation. In regions where the free propagation is
very slow, the step sizes for the constrained propagation are about
two to four orders of magnitude larger, resulting in a much faster integration.     

\begin{figure}[tb]
\includegraphics[width=.9\columnwidth]{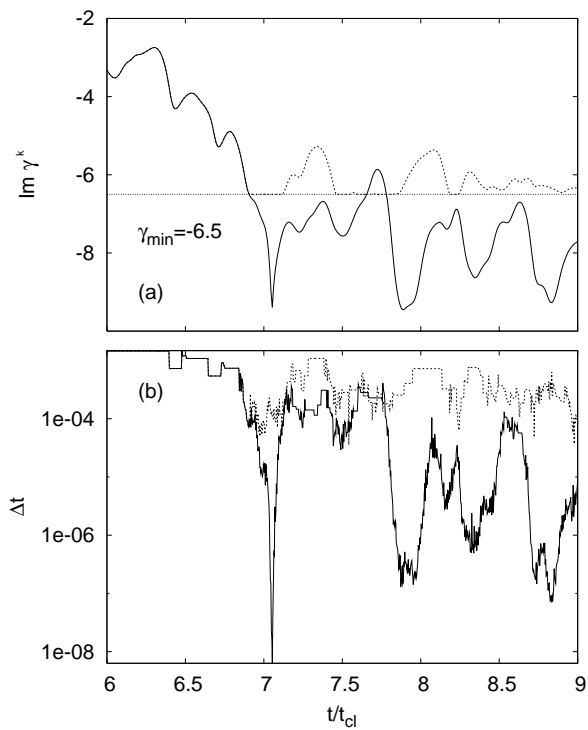}
\caption{Comparison between free and constrained GWP propagation of the same
  initial wave function consisting of a superposition of eight GWP in the
  2D diamagnetic Kepler problem:
(a) 
  Normalization parameter that first reaches $\g_{\rm min}$, (b)
  comparison of the step sizes $\Delta t$ used by the integrator (variable step Adams
  method).  Solid lines
  denote free propagation, dashed lines denote propagation 
  with the constraints $\g^k \geq \g_{\rm min}=-6.5$, $k=1,...,N$. 
 In both calculations the same error tolerances were used.}
\label{fig_div}
\end{figure}

The magnitude of $I$ at its minimum is a measure of the accuracy of the variational
approximation \cite{Raa00,Lub05a}. Therefore a comparison of the minima
$I|_{\dot \z_{\rm abs}}$ 
and
$I|_{\dot \z_{\rm con}}$ allows for an estimate of the loss of accuracy
introduced by the constraints.
A comparison of the minima shows that  $I|_{\dot \z_{\rm con}}$
is slightly increased at
$t\approx 6.9 \, t_{\rm cl}$ with respect to $I|_{\dot \z_{\rm abs}}$
but at later times, one approximation is about as good as the
other in the average, although a poorer approximation of the constrained
wave function to the exact one  would be expected.
However it has been shown that the approximate
wave function determined by TDVP is not always the 'best' possible approximation of
the trial function to the exact wave function \cite{Lub05a}. There might be regions on
the manifold of the trial function that are closer to the exact wave
function than the function determined variationally, especially when the manifold
has a large curvature and long time intervals are considered.
This fact, together with the insensitivity of the wave function to small
variations of the parameters in some directions in case of a singular
matrix, may explain the behavior of only temporary slight loss of
accuracy introduced by the constraints.  
The insensitivity of the trial wave function to the constraints 
can also be deduced from the auto-correlation functions $C(t)= \langle
\chi(t=0)|\chi(t)\rangle$ obtained by both methods since they
almost coincide and no deviation from
each other could be seen in any figure.  

\begin{figure}[tb]
\includegraphics[width=\columnwidth]{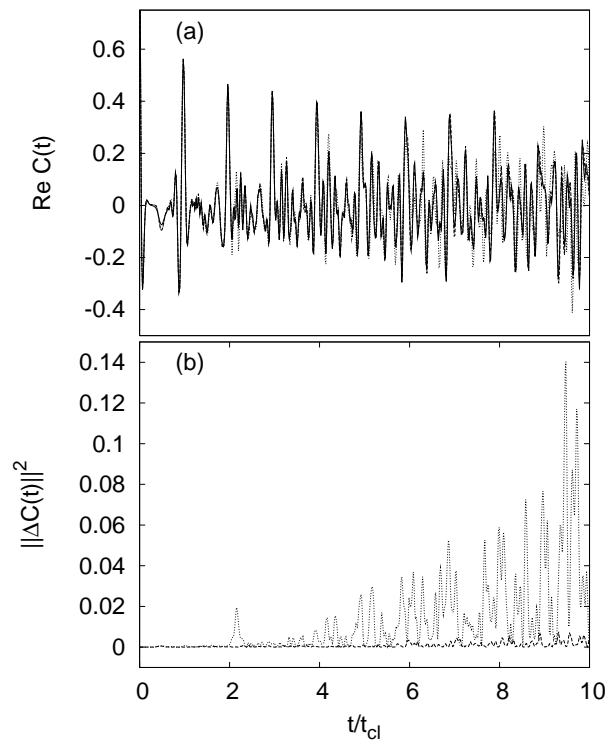}
\caption{(a) Real part of the auto-correlation function. The initial wave
  function for the 2D diamagnetic Kepler problem is a superposition of 20 GWP. Variational propagation with the
  constraints $\g^k\geq -6.5$ (dashed line) is compared with numerically exact
  calculations (solid line) and with frozen GWP propagation (dotted line).
  The results of the constrained calculation and the numerically exact
  calculation practically coincide, and nearly no deviation is visible.
  (b) Deviations of the auto-correlation functions calculated by constrained
  GWP (dashed line) and by frozen GWP (dotted line) from the exact one.} \label{Ct}
\end{figure}

To demonstrate the accuracy of the constrained GWP method a superposition of 20
GWP having all the same width and zero momenta, equally distributed on an
equidistant grid was used as the initial wave packet. This initial wave
packet was propagated by three different methods. 
The real parts of the resulting auto-correlation functions are plotted in figure \ref{Ct}(a).
The imaginary parts, not shown in a the figure, exhibit similar behavior.
For reference the
numerically exact propagation was performed by the split operator method
\cite{Fei82} (solid line). The result of our constrained ($\g^k\ge \-6.5$) GWP propagation
(dashed line) is mostly  very accurate and nearly no deviation from the
exact solution is visible for many classical periods. 
By contrast, the result obtained from a frozen
Gaussian propagation (dotted line) turns out to be much more inaccurate.
This becomes particularly apparent in figure \ref{Ct}(b), where the deviation between the
exact time signal and the time
signals obtained from constrained (dashed) and from
frozen width (dotted) propagation is plotted. 
For short times both methods are very accurate and nearly no
deviation between the time signals is visible. With increasing time, however,
the accuracy of the frozen width calculation is lost much faster
 than that of the constrained propagation. 
This is not completely unexpected since the constrained trial
function still has more free variational parameters than the frozen GWP method and therefore the
constrained calculation is slower.
Note that an unconstrained propagation of these 20 GWP with variable
widths according to the TDVP would not be possible. 
With the propagated wave packet at hand it is straightforward to obtain e.g.\ the
eigenvalues of the Hamiltonian by Fourier transform or harmonic
inversion (for a review see e.g.\ \cite{Mai99}) of the auto-correlation
function  or to extract eigenfunctions of the system (e.g.\ \cite{Hel86}). 

\section{summary}\label{summary}
A novel method to overcome the matrix singularity problem in the
variational Gaussian wave packet method has been proposed. The method is
based on applying  nonholonomic inequality constraints on the motion of the
GWP. The constraints must be chosen to prevent the matrix from becoming
singular.
From the inequality constrained TDVP a simple matrix equation for the time derivatives
of the parameters is obtained just as in unconstrained TDVP. The method is
in fact 
applicable for arbitrary
trial functions and inequality constraints. 
For the GWP trial functions we found it sufficient in most cases to apply simple bounds
on the normalization parameters 
to regularize the matrix and to obtain well-behaved equations of motion, rendering the
integration orders of magnitude faster. 
The loss of accuracy of the method 
caused by the constraints is found to be negligible for
sufficiently many GWP. 
The method allows for
the propagation of a large number of coupled GWP, as compared to the
unconstrained GWP propagation, and guarantees accurate
results within reasonable time. Our method for time propagation of constrained coupled Gaussian wave
packets presented in this paper will be very powerful in a large variety
of future applications to overcome the problems with ill-conditioned
and stuck differential equations.


\begin{thebibliography}{23}
\expandafter\ifx\csname natexlab\endcsname\relax\def\natexlab#1{#1}\fi
\expandafter\ifx\csname bibnamefont\endcsname\relax
  \def\bibnamefont#1{#1}\fi
\expandafter\ifx\csname bibfnamefont\endcsname\relax
  \def\bibfnamefont#1{#1}\fi
\expandafter\ifx\csname citenamefont\endcsname\relax
  \def\citenamefont#1{#1}\fi
\expandafter\ifx\csname url\endcsname\relax
  \def\url#1{\texttt{#1}}\fi
\expandafter\ifx\csname urlprefix\endcsname\relax\def\urlprefix{URL }\fi
\providecommand{\bibinfo}[2]{#2}
\providecommand{\eprint}[2][]{\url{#2}}

\bibitem[{\citenamefont{Heller}(1975)}]{Hel75}
\bibinfo{author}{\bibfnamefont{E.~J.} \bibnamefont{Heller}},
  \bibinfo{journal}{J.\ Chem.\ Phys.} \textbf{\bibinfo{volume}{62}},
  \bibinfo{pages}{1544} (\bibinfo{year}{1975}).

\bibitem[{\citenamefont{Heller}(1976{\natexlab{a}})}]{Hel76}
\bibinfo{author}{\bibfnamefont{E.~J.} \bibnamefont{Heller}},
  \bibinfo{journal}{J.\ Chem.\ Phys.} \textbf{\bibinfo{volume}{64}},
  \bibinfo{pages}{63} (\bibinfo{year}{1976}{\natexlab{a}}).

\bibitem[{\citenamefont{Sawada et~al.}(1985)\citenamefont{Sawada, Heather,
  Jackson, and Metiu}}]{Saw85}
\bibinfo{author}{\bibfnamefont{S.-I.} \bibnamefont{Sawada}},
  \bibinfo{author}{\bibfnamefont{R.}~\bibnamefont{Heather}},
  \bibinfo{author}{\bibfnamefont{B.}~\bibnamefont{Jackson}}, \bibnamefont{and}
  \bibinfo{author}{\bibfnamefont{H.}~\bibnamefont{Metiu}},
  \bibinfo{journal}{J.\ Chem.\ Phys.} \textbf{\bibinfo{volume}{83}},
  \bibinfo{pages}{3009} (\bibinfo{year}{1985}).

\bibitem[{\citenamefont{Heather and Metiu}(1986)}]{Hea86}
\bibinfo{author}{\bibfnamefont{R.}~\bibnamefont{Heather}} \bibnamefont{and}
  \bibinfo{author}{\bibfnamefont{H.}~\bibnamefont{Metiu}},
  \bibinfo{journal}{J.\ Chem.\ Phys.} \textbf{\bibinfo{volume}{84}},
  \bibinfo{pages}{3250} (\bibinfo{year}{1986}).

\bibitem[{\citenamefont{Hansen et~al.}(1989)\citenamefont{Hansen, Henriksen,
  and Billing}}]{Han89}
\bibinfo{author}{\bibfnamefont{F.}~\bibnamefont{Hansen}},
  \bibinfo{author}{\bibfnamefont{N.~E.} \bibnamefont{Henriksen}},
  \bibnamefont{and} \bibinfo{author}{\bibfnamefont{G.~D.}
  \bibnamefont{Billing}}, \bibinfo{journal}{J.\ Chem.\ Phys.}
  \textbf{\bibinfo{volume}{90}}, \bibinfo{pages}{3060} (\bibinfo{year}{1989}).

\bibitem[{\citenamefont{Skodje and Truhlar}(1984)}]{Sko84}
\bibinfo{author}{\bibfnamefont{R.~T.} \bibnamefont{Skodje}} \bibnamefont{and}
  \bibinfo{author}{\bibfnamefont{D.~G.} \bibnamefont{Truhlar}},
  \bibinfo{journal}{J.\ Chem.\ Phys.} \textbf{\bibinfo{volume}{80}},
  \bibinfo{pages}{3123} (\bibinfo{year}{1984}).

\bibitem[{\citenamefont{Kay}(1989)}]{Kay89}
\bibinfo{author}{\bibfnamefont{K.~G.} \bibnamefont{Kay}},
  \bibinfo{journal}{Chem. Phys.} \textbf{\bibinfo{volume}{137}},
  \bibinfo{pages}{165} (\bibinfo{year}{1989}).

\bibitem[{\citenamefont{Fab\v{c}i\v{c}
  et~al.}(2007)\citenamefont{Fab\v{c}i\v{c}, Main, and Wunner}}]{Fab07}
\bibinfo{author}{\bibfnamefont{T.}~\bibnamefont{Fab\v{c}i\v{c}}},
  \bibinfo{author}{\bibfnamefont{J.}~\bibnamefont{Main}}, \bibnamefont{and}
  \bibinfo{author}{\bibfnamefont{G.}~\bibnamefont{Wunner}},
  \bibinfo{journal}{Nonlinear Phenomena in Complex Systems}
  \textbf{\bibinfo{volume}{10}}, \bibinfo{pages}{86} (\bibinfo{year}{2007}).

\bibitem[{\citenamefont{Horenko et~al.}(2004)\citenamefont{Horenko, Weiser,
  Schmidt, and Sch\"utte}}]{Hor04}
\bibinfo{author}{\bibfnamefont{I.}~\bibnamefont{Horenko}},
  \bibinfo{author}{\bibfnamefont{M.}~\bibnamefont{Weiser}},
  \bibinfo{author}{\bibfnamefont{B.}~\bibnamefont{Schmidt}}, \bibnamefont{and}
  \bibinfo{author}{\bibfnamefont{C.}~\bibnamefont{Sch\"utte}},
  \bibinfo{journal}{J.\ Chem.\ Phys.} \textbf{\bibinfo{volume}{120}},
  \bibinfo{pages}{8913} (\bibinfo{year}{2004}).

\bibitem[{\citenamefont{Heller}(1981)}]{Hel81}
\bibinfo{author}{\bibfnamefont{E.~J.} \bibnamefont{Heller}},
  \bibinfo{journal}{J.\ Chem.\ Phys.} \textbf{\bibinfo{volume}{75}},
  \bibinfo{pages}{2923} (\bibinfo{year}{1981}).

\bibitem[{\citenamefont{Zoppe et~al.}(2005)\citenamefont{Zoppe, Parkinson, and
  Messina}}]{Zop05}
\bibinfo{author}{\bibfnamefont{J.}~\bibnamefont{Zoppe}},
  \bibinfo{author}{\bibfnamefont{M.~L.} \bibnamefont{Parkinson}},
  \bibnamefont{and} \bibinfo{author}{\bibfnamefont{M.}~\bibnamefont{Messina}},
  \bibinfo{journal}{Chem.\ Phys.\ Lett.} \textbf{\bibinfo{volume}{407}},
  \bibinfo{pages}{308} (\bibinfo{year}{2005}).

\bibitem[{\citenamefont{Dirac}(1930)}]{Dir30}
\bibinfo{author}{\bibfnamefont{P.~A.~M.} \bibnamefont{Dirac}},
  \bibinfo{journal}{Proc. Cam. Phil. Soc.} \textbf{\bibinfo{volume}{26}},
  \bibinfo{pages}{376} (\bibinfo{year}{1930}).

\bibitem[{\citenamefont{Frenkel}(1934)}]{Fre34}
\bibinfo{author}{\bibfnamefont{J.}~\bibnamefont{Frenkel}},
  \emph{\bibinfo{title}{Wave mechanics, advanced general theory}}
  (\bibinfo{publisher}{Clarendon Press}, \bibinfo{address}{Oxford},
  \bibinfo{year}{1934}).

\bibitem[{\citenamefont{McLachlan}(1964)}]{McL64}
\bibinfo{author}{\bibfnamefont{A.~D.} \bibnamefont{McLachlan}},
  \bibinfo{journal}{Mol. Phys.} \textbf{\bibinfo{volume}{8}},
  \bibinfo{pages}{39} (\bibinfo{year}{1964}).

\bibitem[{\citenamefont{Kramer and Saraceno}(1983)}]{Kra83}
\bibinfo{author}{\bibfnamefont{P.}~\bibnamefont{Kramer}} \bibnamefont{and}
  \bibinfo{author}{\bibfnamefont{M.}~\bibnamefont{Saraceno}},
  \emph{\bibinfo{title}{Geometry of the time-dependent variational principle in
  quantum mechanics}} (\bibinfo{publisher}{Berlin: Springer},
  \bibinfo{address}{Lecture notes in physics}, \bibinfo{year}{1983}).

\bibitem[{\citenamefont{Heller}(1976{\natexlab{b}})}]{Hel76a}
\bibinfo{author}{\bibfnamefont{E.~J.} \bibnamefont{Heller}},
  \bibinfo{journal}{J.\ Chem.\ Phys.} \textbf{\bibinfo{volume}{65}},
  \bibinfo{pages}{4979} (\bibinfo{year}{1976}{\natexlab{b}}).

\bibitem[{\citenamefont{Friedrich and Wintgen}(1989)}]{Fri89}
\bibinfo{author}{\bibfnamefont{H.}~\bibnamefont{Friedrich}} \bibnamefont{and}
  \bibinfo{author}{\bibfnamefont{D.}~\bibnamefont{Wintgen}},
  \bibinfo{journal}{Phys. Rep.} \textbf{\bibinfo{volume}{183}},
  \bibinfo{pages}{37} (\bibinfo{year}{1989}).

\bibitem[{\citenamefont{Hasegawa et~al.}(1989)\citenamefont{Hasegawa, Robnik,
  and Wunner}}]{Has89}
\bibinfo{author}{\bibfnamefont{H.}~\bibnamefont{Hasegawa}},
  \bibinfo{author}{\bibfnamefont{M.}~\bibnamefont{Robnik}}, \bibnamefont{and}
  \bibinfo{author}{\bibfnamefont{G.}~\bibnamefont{Wunner}},
  \bibinfo{journal}{Prog. Theor. Phys. Suppl.} \textbf{\bibinfo{volume}{98}},
  \bibinfo{pages}{198} (\bibinfo{year}{1989}).

\bibitem[{\citenamefont{Raab}(2000)}]{Raa00}
\bibinfo{author}{\bibfnamefont{A.}~\bibnamefont{Raab}},
  \bibinfo{journal}{Chem.\ Phys.\ Lett.} \textbf{\bibinfo{volume}{319}},
  \bibinfo{pages}{674} (\bibinfo{year}{2000}).

\bibitem[{\citenamefont{Lubich}(2005)}]{Lub05a}
\bibinfo{author}{\bibfnamefont{C.}~\bibnamefont{Lubich}},
  \bibinfo{journal}{Math. Comp.} \textbf{\bibinfo{volume}{74}},
  \bibinfo{pages}{765} (\bibinfo{year}{2005}).

\bibitem[{\citenamefont{Feit et~al.}(1982)\citenamefont{Feit, Fleck, and
  Steiger}}]{Fei82}
\bibinfo{author}{\bibfnamefont{M.~D.} \bibnamefont{Feit}},
  \bibinfo{author}{\bibfnamefont{J.~A.} \bibnamefont{Fleck},
  \bibfnamefont{Jr.}}, \bibnamefont{and}
  \bibinfo{author}{\bibfnamefont{A.}~\bibnamefont{Steiger}},
  \bibinfo{journal}{J. Comp. Phys.} \textbf{\bibinfo{volume}{47}},
  \bibinfo{pages}{412} (\bibinfo{year}{1982}).

\bibitem[{\citenamefont{Main}(1999)}]{Mai99}
\bibinfo{author}{\bibfnamefont{J.}~\bibnamefont{Main}}, \bibinfo{journal}{Phys.
  Rep.} \textbf{\bibinfo{volume}{316}}, \bibinfo{pages}{233}
  (\bibinfo{year}{1999}).

\bibitem[{\citenamefont{Reimers and Heller}(1986)}]{Hel86}
\bibinfo{author}{\bibfnamefont{J.~R.} \bibnamefont{Reimers}} \bibnamefont{and}
  \bibinfo{author}{\bibfnamefont{E.~J.} \bibnamefont{Heller}},
  \bibinfo{journal}{J. Phys. A} \textbf{\bibinfo{volume}{19}},
  \bibinfo{pages}{2559} (\bibinfo{year}{1986}).

\end{thebibliography}

\end{document}